# THE ORIGIN OF THE E/Z ISOMER RATIO OF IMINES IN THE INTERSTELLAR MEDIUM

Juan García de la Concepción[1,2], Izaskun Jiménez-Serra[2], José Carlos Corchado[3], Víctor M. Rivilla[2,4], and Jesús Martín-Pintado[2]

## Abstract

Recent astronomical observations of both isomers E and Z of imines such as cyanomethanimine, ethanimine and 2-propyn-1-imine, have revealed that the abundances in the ISM of these isomers differ by factors of ~3-10. Several theories have been proposed to explain the observed behavior, but none of them successfully explains the [E]/[Z] ratios. In this work we present a detailed study of the kinetics of the one-step E-Z isomerization reactions of cyanomethanimine, ethanimine and 2-propyn-1-imine under interstellar conditions (in the 10-400 K temperature range). This reaction was previously thought to be non-viable in the ISM due to its associated high-energy barrier (about 13,000 K). In this Letter, we show that considering the multidimensional small curvature tunneling approximation, the tunneling effect enables the isomerization even at low temperatures. This is due to the fact that the representative tunneling energy lies in the vibrational ground state of the least stable isomer up to approximately 150 K, making the reaction constants of the isomerization from the least stable to the most stable isomer basically constant. The predicted [E]/[Z] ratios are almost the same as those reported from the astronomical observations for all imines observed. This study demonstrates that the [E]/[Z] ratio of imines in the ISM strongly depends on their relative stability.

---

[1]Corresponding author jgarcia@cab.inta-csic.es

[2] Centro de Astrobiología (CSIC-INTA), Ctra. de Ajalvir Km. 4, Torrejón de Ardoz, 28850 Madrid, Spain

[3] Departamento de Ingeniería Química y Química Física, Facultad de Ciencias, and ICCAEx. Universidad Extremadura. Badajoz, Spain

[4] INAF-Osservatorio Astrofisico di Arcetri Largo E. Fermi, 5 I-50125, Firenze, Italy

# 1. INTRODUCTION

Imines are key intermediates in the formation of biologically important precursors such as nitrogenous bases and amino acids, which are the main building blocks of nucleic acids and proteins (Woon 2002). Imines present two different isomers, [E] and [Z], depending on the orientation of the hydrogen atom of the –NH group. Observations in the interstellar medium (ISM) of the [E] and [Z] isomers of imines such as cyanomethanimine, ethanimine, or 2-propyn-1-imine (hereafter propynimine), have shown different behaviors for the measured [E]/[Z] abundance ratios: [E]/[Z]~3 ethanimine (Loomis et al. 2013), [Z]/[E]~6 for cyanomethanimine (Rivilla et al. 2019), and [Z]/[E]>2 for propynimine (Bizzocchi. et al. 2020). Due to the high energy barrier for isomerization, numerous efforts have been made for explaining the observed [*E*]/[*Z*] ratios of imines, mainly based on their formation and/or destruction reactions. However, none of the proposed mechanisms can either explain the observations or do it consistently for all imines observed (Balucani et al., 2018; Baiano. et al., 2020; Bizzocchi et al. 2020; Lupi et al., 2020; Melli et al. 2018; Puzzarini 2015; Quan et al. 2016; Shingledecker et al., 2020).

Z to E (and vice-versa) isomerization of some imines through nitrogen inversion reactions has been previously studied, as well as their relative stabilities and their formation (Balucani et al. 2018; Baiano et al., 2020; Vazart et al., 2015). These studies concluded that this transformation is unlikely, due to the high-energy barrier associated with this isomerization. In addition, the imaginary frequency of this process has been determined to be relatively low (of the order of 1100i cm$^{-1}$, Vazart et. al 2015a; Puzzarini 2015), which could suggest that quantum tunneling effect is not significant. However, quantum tunneling is a very important process at very low temperatures and in enzymatic reactions (Truhlar 2006), where the energy of the system is not high enough to overcome the energy barrier of the reaction, but the implied particle still has a finite probability to pass through the barrier.

The contribution of this quantum effect to the nitrogen inversion reaction has been previously approximated with a monodimensional treatment that neglects the reaction path curvature (Bao & Truhlar 2017). Curvature effects can increase notably the importance of tunneling, especially at low temperatures. Reaction path curvature in a multidimensional system appears as a consequence of the coupling of the motion to the remaining degrees of freedom along the reaction coordinate. Because of the curvature, the tunneling paths do not lie under the minimum energy path (MEP) but, as a result of centrifugal forces, the system slides to the concave side of the reaction valley. This enables the tunneling path to "cut a corner", increasing the tunneling effect with respect to a monodimensional treatment (Bao & Truhlar 2017).

In this work we have carried out the first detailed study of the isomerization of cyanomethanimine, ethanimine and propynimine through the nitrogen inversion reaction for interstellar conditions using a multidimentional treatment of quantum tunneling (Bao & Truhlar 2017 and references therein). Our results fully explain the observed [E]/[Z] abundance ratios of the imines reported in the ISM, and demonstrate that their final abundances are dominated by isomerization reactions. We also evaluate the kinetic isotope effects (KIE) in this isomerization by substituting NH by ND in the imines. Deuterated imines, which would form mostly on grains, could give additional information about the grain chemistry of the imines.

## 2. COMPUTATIONAL DETAILS

### 2.1. *Electronic structure calculations*

The isomerization reaction involves the C1-N-H bending mode of the imine, which connects the E and Z isomers through a transition structure (TS) where the C1-N-H angle is near 180⁰ for the three imines considered in this work (Figure 1).

All stationary points [namely, the E and Z isomers and the saddle points (SP) that link the minima] have been optimized at the ab initio Møller–Plesset perturbation theory MP2 (Head-Gordon et al, 1988) in combination with the aug-cc-pVTZ (Dunning Jr. 1989; Kendall. et al. 1992) basis set employing the Gaussian 16 software package (Frisch et al. 2016). Frequency calculations were computed at the same level of theory obtaining zero and one imaginary frequency for the minima and the saddle points respectively. In order to give more accurate results we corrected the electronic energies of the MP2/aug-cc-pVTZ-optimized geometries with the coupled cluster theory for single and double excitations augmented with perturbative triples (Raghavachari et al.1989). We also included correlated basis functions with the F12 correction (Pavosevic et al. 2016) and resolution-of-identity approach (RI) for the correlation integrals (Feyereisen et al. 1993) [CCSD(T)-F12/RI]. The inclusion of this explicit electronic correlation is a good alternative to the basis set extrapolation since it is known to give results near the basis set limit with smaller basis sets. To this purpose we used the cc-pVTZ-F12 as special orbital basis set and a near complete auxiliary basis set cc-pVTZ-F12-CABS. For the coulomb and correlation fitting the aug-cc-pVTZ basis set was used. This method gives a realistic estimation for single configuration functions as the cases we treat here, since the systems are pure closed shell singlets. For the coupled cluster calculations we used the ORCA software (Neese 2012; Neese et al. 2020). For the visualization of the molecular geometries we used the *Cylview* program (Legault 2010).

### 2.2. *Kinetic methods*

For computing the reaction rate constants, we used the canonical variational transition state theory (CVT) with the multidimensional small curvature tunneling approximation (SCT). For the sake of comparison, we also computed rate constants using the widely employed, standard method of conventional (i.e. non-variational) transition state theory with one-dimensional tunneling through Eckart function corrections.

For the canonical variational transition state theory (CVT) (Bao & Truhlar 2017), the reaction rate constants have been calculated in the temperature range of 10-400 K using the Pilgrim software (Ferro-Costas et al. 2020). The unimolecular rate coefficient is given by

$$K^{CVT/SCT}(T) = \kappa^{SCT} \Gamma^{CVT} \frac{K_B T}{h} \frac{Q^{VT}}{Q^R} e^{(\frac{-E^{VT}}{K_B T})} \qquad (1)$$

where $K_B$ is the Boltzmann constant, $h$ is the Planck constant, $\kappa^{SCT}$ is the small curvature multidimensional tunneling transmission coefficient and $\Gamma^{CVT}$ is the canonical variational transition state recrossing coefficient, which is the relationship between $K^{CVT}$ and the rate coefficient of the conventional transition state theory $K^{TST}$. $Q^{VT}$ and $Q^{R}$ are the total partition functions of the variational transition state and the reactant, respectively, and $E^{VT}$ is the potential energy of the variational transition state.

Dual level calculations were carried out to compute $K^{CVT}$ in order to correct the MEP energies previously computed at MP2/aug-cc-pVTZ using the interpolated single-point energies (ISPE) algorithm (Chuang et al. 1999). For this purpose we computed single point energies on the reactant, product, saddle point and six points along the MEP at the coupled cluster method mentioned above.

As we are working at very low temperatures the SCT was computed with the quantized-reactant-state tunneling (QRST) approximation (Wonchoba et al. 1994; Wonchoba et al. 1995), since at low temperatures molecules on the reaction path are better described on discrete energy levels instead of a continuous energy level as is commonly assumed in most tunneling approximations.

In the computation of conventional TST rate coefficients, the $\kappa^{SCT}$ was substituted by the one-dimensional tunneling transmission coefficient for an Eckart polynomial function fitted to the reactants, products, and saddle point energies, $\kappa^{Eckart}$, (Johnson & Heickler 1962). In addition, the recrossing coefficient $\Gamma^{CVT}$ was set to one and $Q^{VT}$ and $E^{VT}$ were replaced by the total partition functions and potential energy of the saddle point, respectively, as follows:

$$K^{TST/Eckart}(T) = \kappa^{Eckart} \frac{K_B T}{h} \frac{Q^{SP}}{Q^R} e^{\left(\frac{-E^{SP}}{K_B T}\right)} \qquad (2)$$

For the computation of the KIEs, we relate the rate constants of the lighter compound (L) with the heavier isotopologue (H) as

$$KIE = \frac{K_L}{K_H} \qquad (3)$$

Finally, we note that forward pathway (*i.e.*, from $S < 0$ to $S > 0$, being S the reaction coordinate in Bohr) hereafter means the movement along the reaction path from the least stable isomer to the most stable one. The deuterated isotopologues of the imines obtained by substituting NH by ND, are denoted as E(D) and Z(D).

## 3. RESULTS

The energetic data obtained with the above-mentioned methodology lie close to previous high-level results as inferred from Table 1, which shows the relative vibrationally adiabatic energies (i.e. zero-point inclusive potential energies) of the stationary points with respect to the least stable isomer (*E* for cyanomethanimine and propynimine, and *Z* for ethanimine).

For the three imines, the ZPE (zero point energy) does not change notably along the MEP since the frequencies remain almost constant during the reaction coordinate. This effect indicates that variational effects should be very small. In fact, the canonical variational transition state recrossing coefficient $\Gamma^{CVT}$ is 1 across the whole temperature range, which means that these reactions can be well described within the transition state theory (TST), since transition-state recrossing is negligible. These results point out that the notable differences between the rate constants obtained with CVT/SCT and TST, especially below 200 K, are due to the quantum tunneling treatments (see below).

In Figure 2, we plot Log $K$(T) against T(K) for the isomerization reactions of these imines for the two methods employed in this study (TST/Eckart and CVT/SCT). Table 2 reports the derived values of Log $K$(T) for the forward and backward isomerization reactions for the three imines for the method that best reproduces the observations (CVT/SCT; see below). This Table also reports the [Z]/[E] ratios of the reactions rates for cyanomethanimine and propynimine, and the [E]/[Z] ratio of the rates for ethanimine.

From Figure 2, we find that the CVT/SCT calculations are the ones that best explain the behavior of the [E]/[Z] abundance ratios for the observed imines in the ISM. The results from the conventional TST with the one-dimensional Eckart tunneling correction provide very low reaction rates constants, especially below 200 K, falling well below from what would be observable in the ISM. The poor agreement obtained with the TST/Eckart are due to the fact that we have to approximate the shape of the potential energy to an Eckart potential in order to obtain the tunneling probabilities. The parameters obtained with this model are used to fit the reaction enthalpy at 0 K. In addition, the magnitude of the second derivative of the energy should be approximated as $4\pi^2 \mu v^{\ddagger 2}$; where $v^{\ddagger}$ is the magnitude of the imaginary frequency of the saddle point and $\mu$ the reduced mass. However, in some cases these approximations may not be accurate enough to give quantitative results, as in the case at hand. In any case, the shape of the potential is not the only reason for the poor performance of the Eckart transmission coefficient.

The results obtained with a multidimensional treatment of the tunneling effect (CVT/SCT) show that the forward and backward isomerizations are controlled by tunneling from 10 to ~340 K. In the forward direction, at 10 K the rate constants with SCT corrections ($K^{CVT/SCT}$) are 550 orders of magnitude higher than without taking into account tunneling effects ($K^{CVT}$). Note that such large differences between the rate constant corrected with SCT with respect to the one without SCT, have also been found for other systems at low temperatures (see the case of the isomerization of carbenes, with a difference of 152 orders of magnitude at 8 K; Zuev et al. 2003). These results at very low temperatures suggest that the reaction proceeds without activation energy. In this case the isomerization occurs exclusively from the vibrational ground state of the highest energy isomer, at much lower energy than the activation barrier. This happens up to relatively high temperatures, where the representative tunneling energies are very close to $v = 0$. This effect is reflected in the plots of Figure 2 where the rate constants show a plateau up to these temperatures. After ~150 K there is a transition from the ground-state tunneling to the thermally activated reaction, which can be inferred from the appearance of a slope in the plots. On the other hand, the representative tunneling energy at very low temperatures lie above the vibrational ground state of the least stable isomer; tunneling through the barrier in the endothermic direction requires an energy that is hard to obtain at very low temperatures.

In this sense, the higher the energy difference between the isomers, the lower rate for the back isomerization.

The equilibrium constants computed as the ratios $K^{CVT/SCT}$[E·to Z]/$K^{CVT/SCT}$[Z to E] for cyanomethanimine and propynimine and $K^{CVT/SCT}$[Z to E]/$K^{CVT/SCT}$[E to Z] for ethanimine (collected in Table 2) show that at low temperatures almost all the cyanomethanimine and propynimine will be as the Z isomer, whereas for ethanimine the equilibrium will be almost totally displaced to the E isomer. Note that, as discussed in Section 4, the [E]/[Z] ratio will depend on the energy difference between the isomers, as hinted by Rivilla et al. (2019), but we can explain this ratio from the relationship of the kinetic rate constants.

The isotopic substitution (with NH replaced by ND) has a notable influence in the kinetics of the isomerization for the three imines. The imaginary frequency of the saddle points decreases to 861.1i, 884.7i and 881.9i cm$^{-1}$ for deuterated cyanomethanimine, ethanimine and propynimine, respectively. This effect widens the MEP and consequently the vibrationally adiabatic potential, decreasing in turn the quantum tunneling effects. This is the expected behavior for a tunneling atom with a mass a factor of two higher. Figure 2 indeed shows that the rate constants of the deuterated imines decrease almost ten orders of magnitude at low temperatures with respect to their H-analogous, and remain almost constant in the plateau region of the plots. When the reactions become thermally-controlled at temperatures >340 K, this difference becomes negligible. This change of the KIE with temperature can also be seen in Figure 2 for cyanomethanimine (note that this behavior is almost the same for ethanimine and propynimine).

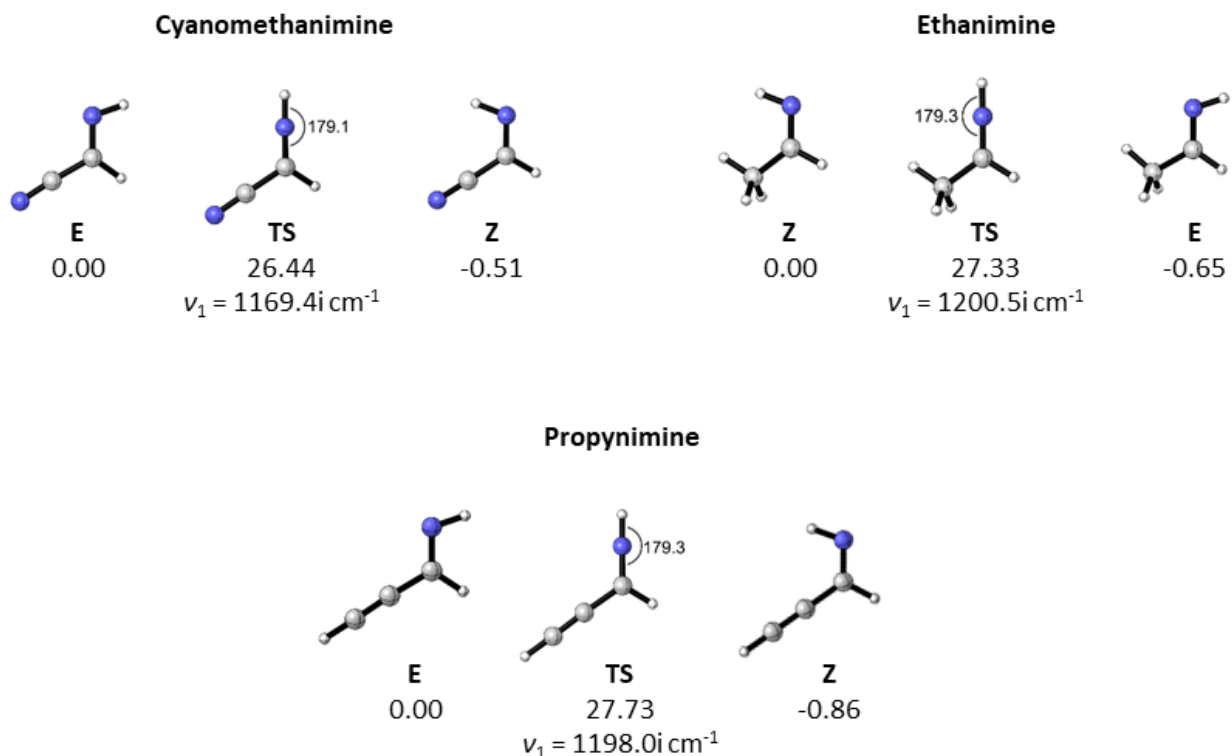

Figure 1. Optimized geometries of the E, Z and TS stationary points for the three imines considered in this work, and its relative vibrationally adiabatic energies with respect to the less stable isomer (E for cyanomethanimine and propynimine, and Z for ethanimine) in units of kcal/mol.

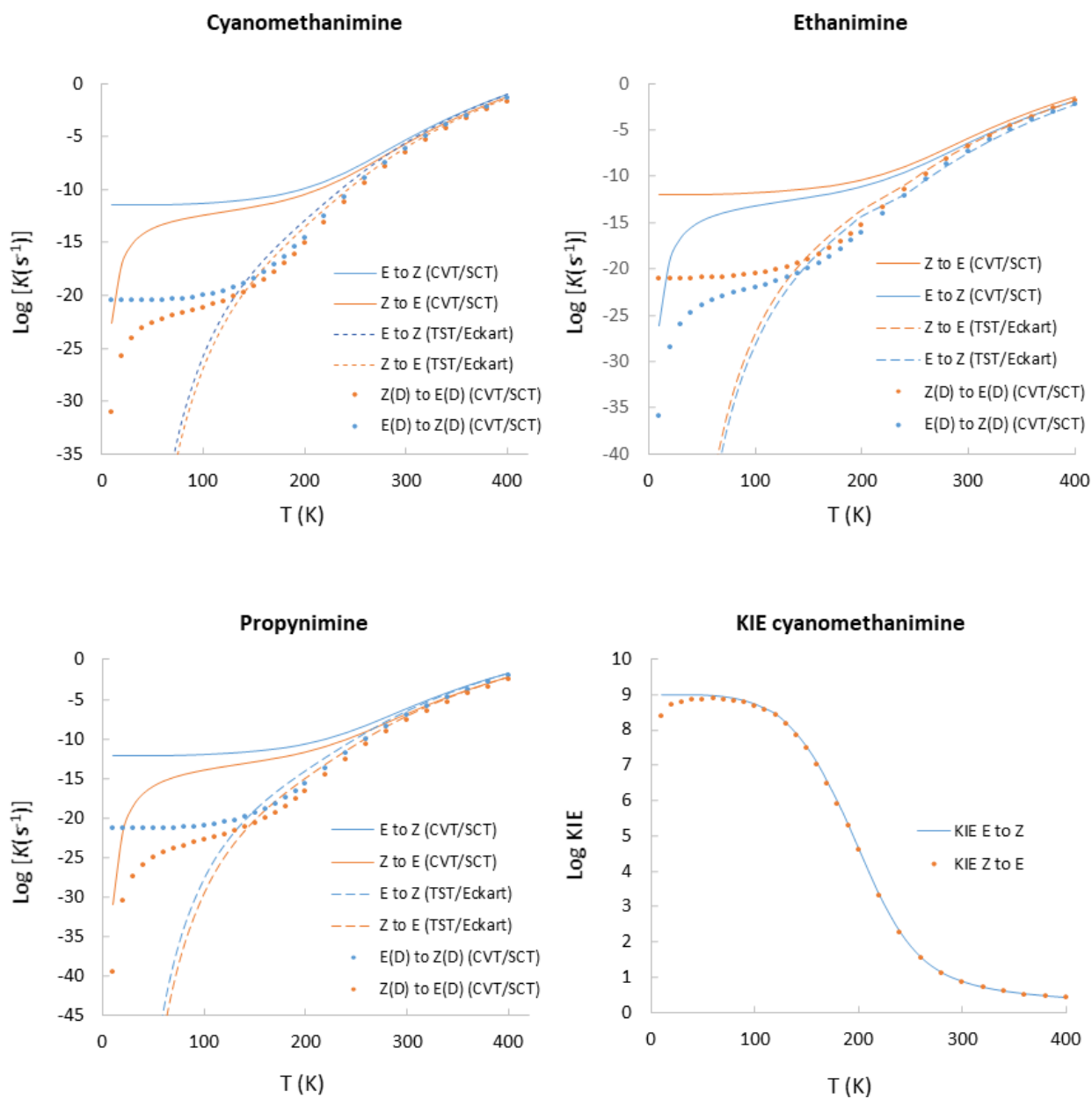

Figure 2. Plots of Log K(T) against T(K) for the E to Z isomerization (blue lines) and Z to E (orange lines) with the CVT/SCT (solid lines) and TST/Eckart (dashed lines). Dotted curves are Log K(T) against T(K) for the deuterated isotopologes of the imines (with NH replaced by ND). As an example, a plot of the KIE for cyanomethanimine against T (K) is also shown.

Table 1. Vibrationally adiabatic energies for the E and Z isomers and the transition structure (TS) that connects them. Relative energies are given with respect to the less stable isomer in units of kcal/mol (values is brackets are in Kelvin).

| Stationary points | Cyanomethanimine | | | | Ethanimine | | | | | Propynimine | | |
|---|---|---|---|---|---|---|---|---|---|---|---|---|
| | This work | Shingledecker et al. 2020 | Puzzarini 2015 | Vazart et al. 2015 | This work | Baiano et al. 2020 | Melli et al. 2018 | Quan et al. 2016 | Balucani, et al. 2018 | This work | Bizzocchi et al. 2020 | Lupi et al. 2020 |
| E | 0 | 0 | 0 | 0 | -0.65 (-327.09) | -0.68 (-342.19) | -0.66 (-332.13) | -1.1 (-553.50) | -0.69 (-347.22) | 0 | 0 | 0 |
| Z | -0.51 (-256.64) | -0.52 (-261.68) | -0.47 (-236.51) | -0.55 (-276.77) | 0 | 0 | 0 | 0 | 0 | -0.86 (-432.77) | -0.85 (-427.74) | -0.81 (-407.61) |
| TS | 26.44 (13,305.20) | ----- | 26.03 (13,098.90) | 24.85 (12,505.10) | 27.43 (13,803.40) | 26.97 (13,571.90) | 27.68 (-13929.20) | 32.03 (16,118.20) | 28.18 (14,180.80) | 27.73 (13,753.10) | ----- | ----- |

Table 2. CVT/SCT reaction rate constants (in $s^{-1}$) for the forward and backward isomerization reaction of cyanomethanimine, ethanimine and propynimine and forward/backward ratios.

| T (K) | Cyanomethanimine | | | Ethanimine | | | Propynimine | | |
|---|---|---|---|---|---|---|---|---|---|
| | $K^{CVT/SCT}[\text{E to Z}]$ | $K^{CVT/SCT}[\text{Z to E}]$ | $\dfrac{K^{CVT/SCT}[\text{E to Z}]}{K^{CVT/SCT}[\text{Z to E}]}$ | $K^{CVT/SCT}[\text{E to Z}]$ | $K^{CVT/SCT}[\text{Z to E}]$ | $\dfrac{K^{CVT/SCT}[\text{E to Z}]}{K^{CVT/SCT}[\text{Z to E}]}$ | $K^{CVT/SCT}[\text{E to Z}]$ | $K^{CVT/SCT}[\text{Z to E}]$ | $\dfrac{K^{CVT/SCT}[\text{E to Z}]}{K^{CVT/SCT}[\text{Z to E}]}$ |
| 10 | 3.22E-12 | 2.03E-23 | 1.59E+11 | 9.38E-13 | 6.73E-27 | 1.39E+14 | 6.38E-13 | 1.13E-31 | 5.65E+18 |
| 20 | 3.22E-12 | 7.86E-18 | 4.10E+05 | 9.38E-13 | 8.07E-20 | 1.16E+07 | 6.38E-13 | 2.61E-22 | 2.45E+09 |
| 30 | 3.22E-12 | 5.73E-16 | 5.63E+03 | 9.39E-13 | 1.85E-17 | 5.08E+04 | 6.38E-13 | 3.45E-19 | 1.85E+06 |
| 40 | 3.23E-12 | 4.90E-15 | 6.60E+02 | 9.49E-13 | 2.83E-16 | 3.36E+03 | 6.40E-13 | 1.26E-17 | 5.09E+04 |
| 50 | 3.26E-12 | 1.79E-14 | 1.82E+02 | 9.75E-13 | 1.48E-15 | 6.57E+02 | 6.50E-13 | 1.10E-16 | 5.90E+03 |
| 60 | 3.35E-12 | 4.32E-14 | 7.74E+01 | 1.02E-12 | 4.62E-15 | 2.21E+02 | 6.70E-13 | 4.78E-16 | 1.40E+03 |
| 70 | 3.49E-12 | 8.32E-14 | 4.20E+01 | 1.10E-12 | 1.08E-14 | 1.02E+02 | 7.05E-13 | 1.40E-15 | 5.03E+02 |
| 80 | 3.72E-12 | 1.40E-13 | 2.65E+01 | 1.20E-12 | 2.12E-14 | 5.67E+01 | 7.57E-13 | 3.25E-15 | 2.33E+02 |
| 90 | 4.06E-12 | 2.19E-13 | 1.86E+01 | 1.34E-12 | 3.73E-14 | 3.60E+01 | 8.31E-13 | 6.49E-15 | 1.28E+02 |
| 100 | 4.54E-12 | 3.25E-13 | 1.40E+01 | 1.53E-12 | 6.13E-14 | 2.50E+01 | 9.33E-13 | 1.18E-14 | 7.94E+01 |
| 110 | 5.20E-12 | 4.69E-13 | 1.11E+01 | 1.79E-12 | 9.64E-14 | 1.86E+01 | 1.07E-12 | 2.00E-14 | 5.37E+01 |
| 120 | 6.13E-12 | 6.70E-13 | 9.14E+00 | 2.14E-12 | 1.48E-13 | 1.45E+01 | 1.26E-12 | 3.26E-14 | 3.88E+01 |
| 130 | 7.44E-12 | 9.59E-13 | 7.76E+00 | 2.62E-12 | 2.23E-13 | 1.17E+01 | 1.53E-12 | 5.20E-14 | 2.94E+01 |
| 140 | 9.36E-12 | 1.39E-12 | 6.74E+00 | 3.31E-12 | 3.38E-13 | 9.79E+00 | 1.91E-12 | 8.23E-14 | 2.32E+01 |
| 150 | 1.23E-11 | 2.06E-12 | 5.97E+00 | 4.33E-12 | 5.18E-13 | 8.37E+00 | 2.48E-12 | 1.31E-13 | 1.89E+01 |
| 160 | 1.69E-11 | 3.15E-12 | 5.37E+00 | 5.92E-12 | 8.11E-13 | 7.30E+00 | 3.37E-12 | 2.13E-13 | 1.58E+01 |
| 170 | 2.30E-11 | 4.71E-12 | 4.89E+00 | 8.01E-12 | 1.24E-12 | 6.47E+00 | 4.83E-12 | 3.57E-13 | 1.35E+01 |
| 180 | 3.69E-11 | 8.20E-12 | 4.50E+00 | 1.25E-11 | 2.15E-12 | 5.81E+00 | 7.04E-12 | 5.99E-13 | 1.17E+01 |
| 190 | 6.34E-11 | 1.52E-11 | 4.18E+00 | 2.08E-11 | 3.94E-12 | 5.27E+00 | 1.18E-11 | 1.13E-12 | 1.04E+01 |
| 200 | 1.18E-10 | 3.02E-11 | 3.91E+00 | 3.72E-11 | 7.70E-12 | 4.83E+00 | 2.12E-11 | 2.29E-12 | 9.25E+00 |
| 220 | 5.41E-10 | 1.56E-10 | 3.48E+00 | 1.55E-10 | 3.72E-11 | 4.16E+00 | 8.96E-11 | 1.18E-11 | 7.60E+00 |
| 240 | 3.70E-09 | 1.17E-09 | 3.16E+00 | 9.50E-10 | 2.59E-10 | 3.67E+00 | 5.60E-10 | 8.67E-11 | 6.46E+00 |
| 260 | 3.61E-08 | 1.24E-08 | 2.91E+00 | 8.44E-09 | 2.55E-09 | 3.31E+00 | 5.08E-09 | 9.03E-10 | 5.63E+00 |
| 280 | 4.28E-07 | 1.58E-07 | 2.71E+00 | 9.62E-08 | 3.19E-08 | 3.02E+00 | 5.90E-08 | 1.18E-08 | 5.00E+00 |
| 300 | 5.06E-06 | 1.98E-06 | 2.55E+00 | 1.16E-06 | 4.14E-07 | 2.79E+00 | 7.21E-07 | 1.60E-07 | 4.51E+00 |
| 320 | 5.25E-05 | 2.17E-05 | 2.42E+00 | 1.26E-05 | 4.82E-06 | 2.61E+00 | 8.01E-06 | 1.94E-06 | 4.12E+00 |
| 340 | 4.58E-04 | 1.98E-04 | 2.31E+00 | 1.16E-04 | 4.73E-05 | 2.45E+00 | 7.56E-05 | 1.99E-05 | 3.81E+00 |
| 360 | 3.32E-03 | 1.50E-03 | 2.21E+00 | 8.96E-04 | 3.86E-04 | 2.32E+00 | 5.95E-04 | 1.68E-04 | 3.55E+00 |
| 380 | 2.03E-02 | 9.53E-03 | 2.13E+00 | 5.82E-03 | 2.63E-03 | 2.22E+00 | 3.94E-03 | 1.18E-03 | 3.33E+00 |
| 400 | 1.06E-01 | 5.15E-02 | 2.06E+00 | 3.22E-02 | 1.52E-02 | 2.12E+00 | 2.22E-02 | 7.05E-03 | 3.14E+00 |

## 4. COMPARISON WITH OBSERVATIONS

Our results from Section 3 not only follow nicely the trends observed for the isomer ratios of the three imines, but they perfectly match the observed [E]/[Z] ratios observed in the ISM (Table 3). Millimeter observations carried out toward the giant molecular cloud G+0.693-0.027 (hereafter G+0.693) located in the Galactic Center (with $T_{kin}$~150 K) reveal that the ratios [Z]/[E] of cyanomethanimine and [E]/[Z] of ethanimine are ~6 and ~10-15, respectively (Rivilla et al. 2019, Loomis et al. 2013, Rivilla et al. in prep.). Our CVT/SCT results give ratios of 6.0 and 8.4, respectively, at 150 K. In addition, the recently reported detection of Z- propynimine toward the same source (Bizzocchi et al. 2020) gives a lower limit > 1.9 for the [Z]/[E] ratio, which is consistent with our predicted [Z]/[E] ratio of 19.0 at 150 K. For ethanimine, the observed [E]/[Z] ratio toward the molecular cloud SgrB2(N) (with $T_{kin}$~300 K) is ~3, perfectly matching our results of 2.8 for the predicted [E]/[Z] ratio.

We note that our derived CVT/SCT results show that it is possible that the imine isomers ratio follow the expected behavior under thermodynamic equilibrium even at low temperatures thanks to tunneling. This is shown in Table 3, where we report both the calculated equilibrium constants between the E and Z isomers (using equation 4 below) and the ratio obtained using equation 5 proposed to explain the isomer ratios by Rivilla et al. (2019, see below), for the temperature range 140-380 K. Both expressions follow the same trend since the free energy difference ($\Delta G_{reacc}$) between the two isomers remains almost constant for the whole range of temperatures.

$$K_{eq} = \frac{K^{CVT/SCT}[E\ to\ Z]}{K^{CVT/SCT}[Z\ to\ E]} = \frac{[Z]}{[E]} = e^{(\frac{-\Delta G_{reac}}{RT})} \quad (4)$$

$$\frac{[Z]}{[E]} = \frac{1}{g} e^{(\frac{-\Delta(E+ZPE)}{T})} \quad (5)$$

We finally note that any dynamical effects such as pumping or preferential excitation of the bending coordinate would increase the rate of the reaction helping the equilibrium, **understood as the equilibration between the E and Z isomers,** to be reached faster.

Table 3. Observed [Z]/[E] (or[E]/[Z]) abundance ratios of the imines considered in this work, and predictions from the CVT/SCT **and equation 5**.

| Source | $T_{kin}$ (K) | Cyanomethanimine [Z]/[E] | | | Ethanamine [E]/[Z] | | | Propynimine [Z]/[E] | | |
|---|---|---|---|---|---|---|---|---|---|---|
| | | Observation | CVT/SCT results | **Equation 5** | Observation | CVT/SCT results | **Equation 5** | Observation | CVT/SCT results | **Equation 5** |
| SgrB2(N) | ~300 | ----- | ----- | **-----** | ~3 | 2.8 | **3.0** | ----- | ----- | **-----** |
| G+0.693 | ~150 | ~6 | 6.0 | **5.5** | ~10-15 | 8.4 | **8.9** | > 1.9 | 19.0 | **18.0** |

## 5. DISCUSSION

Several theories have been proposed to explain the observed [E]/[Z] abundance ratios of the imines studied in the ISM, but none of them fully explains the observations. The most popular scenario considers that the observed [E]/[Z] (or [Z]/[E]) ratios are a consequence of the formation/destruction mechanisms of imines in the ISM (Herbst et al. 2000; Loomis et al. 2015; Shingledecker et al. 2019). For cyanomethanimine, Vazart et al. (2015b) proposed a [Z]/[E] ratio of 1.9 based on its formation from CN and $CH_2NH$, which lies far from the observed [Z]/[E] ratio of ~6 (Rivilla et al. 2019). Shivani et al. (2017) also found a ratio of 0.9 in the gas phase and 1.0 on grain surfaces assuming that cyanomethanimine forms from NCCN and H. Using astrochemical codes, Zhang et al. (2020) estimate a ratio between 1 and 4 for the [Z]/[E] ratio of cyanomethanimine, but the physical conditions assumed for the C-shocks present in the Galactic Center (and in G+0.693 in particular) are not realistic, because they consider cosmic-ray ionization rates that are factors >100 lower than observed in this region of the Galaxy (see Goto et al. 2014; Zeng et al. 2018). Other works achieve a [Z]/[E] ratio of ~3 (Shingledecker et al. 2020) considering the destruction of the isomers via ion-molecule reactions. However, the same argument does not apply to other imines such as propynimine, for which the predicted [Z]/[E] ratio is ~1 while the observed [Z]/[E] ratio is >1.9 (Bizzocchi et al. 2020). Lupi et al. (2020) again propose a gas-phase formation scenario for the origin of the observed [Z]/[E] ratio (of ~1.5), but their predicted ratio lies a bit short with respect to the lower limit measured in the ISM.

Similar studies have also been carried out to explain the [E]/[Z] ratio observed for ethanimine. Initially, astrochemical codes predicted ratios that deviate largely from observations (Quan et al. 2016). Later on, the reaction of the amidogen (NH) and ethyl radical ($CH_3CH_2$) was proposed as a promising formation route for ethanimine with a predicted [E]/[Z] ratio of 1.2 (Balucani et. al 2018). However, the revision of the latter theoretical calculations did not approach the observed values either (predicted value of 1.4 vs. an observed value of 3; Baiano et. al 2020).

In all these previous works, it was assumed that thermodynamic equilibrium between isomers at the kinetic temperature cannot explain the observed ratio due to the large activation energy barriers of the isomerization. However, our results considering a multi-dimensional treatment of the tunneling effect, reveal that the relative abundances of the isomers of the imines observed in the ISM match the values expected from their thermodynamic stabilities (Table 3). Indeed, the [Z]/[E] ratios predicted for cyanomethanimine and propynimine perfectly match those observed toward G+0.693, while the calculated [E]/[Z] ratios for ethanimine are very similar to those measured toward both G+0.693 and SgrB2N (Table 3), reproducing the observations not only for different sources with different kinetic temperatures but also matching the different trends for the [Z]/[E] or [E]/[Z] ratios depending on the imine species.

In summary, our results show that the transformation from the least stable isomer to the more stable one is possible even at very low temperatures thanks to tunneling, allowing the system to reach the thermodynamic equilibrium. This demonstrates that, in some cases, the one-dimensional treatment of the quantum tunneling notably underestimate the reaction rate constants, and that the origin of the E/Z ratio of imines in the ISM depends exclusively on their relative stabilities, with the latter being understood as their difference in energy as a function of temperature.

J.G.d.l.C acknowledges the Spanish State Research Agency (AEI) through project number MDM-2017-0737 Unidad de Excelencia "María de Maeztu"- Centro de Astrobiología (CSIC-INTA). J.C.C. acknowledges the Junta de Extremadura and European Regional Development Fund, Spain, Project No. GR18010. I.J.-S. and J.M.-P. have received partial support from the Spanish FEDER (ESP2017-86582-C4-1-R), and AEI (PID2019-105552RB-C41). V.M.R. has received funding from the Comunidad de Madrid through the Atracción de Talento Investigador (Doctores con experiencia) Grant (COOL: Cosmic Origins Of Life; 2019-T1/TIC-15379). Computational assistance was provided by the Supercomputer facilities of LUSITANIA founded by Cénits and Computaex Foundation.

# REFERENCES

Avalos, M., Babiano, R., Cintas, M., Jiménez, J. L., Palacios & J. C., Barron, L. D. 1998. Chem. Rev. 98, 2391.
Baiano, C., Lupi, J., Tasinato, N., Puzzarini, C., & Barone, V. 2020, Molecules, 25, 2873.
Balucani, N., Skouteris, D., Ceccarelli, C., Codella, C., Falcinelli, S. & Rosi, M. 2018. Mol. Astrophys. 13, 30.
Bao, J. L., Trulhar, D. G. 2017. Chem. Soc. Rev. 46, 7548.
Bizzocchi, L., Prudenzano, D., Rivilla, V. M., et al. 2020, Astron. Astrophys., 640, A98.
Chuang, Y.-Y., Corchado, J. C., Truhlar, D. G. 1999, J. Phys. Chem. A, 103, 1140.
Cronin, J. R., Pizzarello, S. Science. 1997, 5302, 951.
Cronin, J. R., Pizzarello, S. 1999, Adv. Space Rex. 23, 293.
Adv. Space Rex. Vol. 23, No. 2, pp. 293-299, 1999
Dunning Jr., T. H., 1989. *J. Chem. Phys.* 90, 1007.
Ferro-Costas, G., Truhlar, D. G., Fernández-Ramos A., *Pilgrim*-version 2020.2 (University of Minneapolis, Minnesota, MN, and Universidade de Santiago de Compostela, Spain, 2020).
Feyereisen, M., Fitzerald, G. & Komornicki, A. 1993. Chem. Phys. Lett., 208, 359.
Frisch, M. J., Trucks, G., W., Schlegel, H., B., et al. 2016, Gaussian 16 Revision A.03.
Goto, M., Geballe, T. R., Indriolo, N., Yusef-Zadeh, F., Usuda, T., Henning, T., & Oka, T. 2014, ApJ, 786, 15
Head-Gordon, M., Pople, J. A. & Frisch, M. J. 1988. Chem. Phys. Lett., 153, 503.
Herbst, E., Terzieva, R., & Talbi, D. 2000, MNRAS, 311, 869.
Johnson, H. S., Heickler, J. 1962. J. Phys. Chem. 66, 532.
Kauffman, S. A. 2011 *Life.* 1, 34.
Kendall, R. A., Dunning Jr. T. H. & Harrison, R. J. 1992. J. Chem. Phys., 96, 6796.
Legault, C. Y., 2010. Comput. Programs Biomed., 18, 99.
Loomis, R. A., McGuire, B. A., Shingledecker, C., et al. 2015, ApJ, 799, 34.
Loomis, R. A., Zaleski, D. P., Steber, A. L., et al. 2013, ApJL, 765, L9.
Lupi, J., Puzzarini, C. & Barone, V. 2020, Astrophys. Lett., ACCEPTED.
Melli, A.; Melosso, M.; Tasinato, N.; Bosi, G.; Spada, L.; Bloino, J.; Mendolicchio, M.; Dore, L.; Barone, V. & Puzzarini, C. 2018. Astrophys. J. 855, 13.
Neese, F., WIREs:2012. Comput. Mol. Sci. 2, 73 .
Neese, F., Wennmohs, F., Becker, U. & Riplinger C. 2020. J. Chem. Phys. 152, 224108
Pavosevic, F., Pinski, P., Riplinger, C., Neese, F., & Valeev, E. 2016. J. Chem. Phys., 144, 144109.
Pizzarello, S., Schrader, D. L., Monroe, A. A., & Lauretta, D. S. 2012, PNAS. 109, 11949.
Puzzarini, C., 2015. J. Phys. Chem. A. 119, 11614.
Quan, D., Herbst, E., Corby, J.F., Durr, A. & Hassel, G. 2016. Astrophys. J. 824, 129.
Raghavachari, K., Trucks, G. W., Pople, J. A., & Head-Gordon, M. 1989, Chem. Phys. Lett., 157, 479
Ruiz-Mirazo, K., Briones, C. & de la Escosura, Andrés, Chem. 2014. Chem. Rev. 114, 285.
Rimola, A., Sodupe, M., Ugliengo, P., 2011. Rend. Fis. Acc. Lincei, 22, 137.
Rimola, A., Sodupe, M., Ugliengo, P., 2012. Astrophys. J. 754, 24.
Rivilla, V. M., Martín-Pintado, J., Jiménez-Serra, I., et al. 2019, MNRAS, 483, L114

Shingledecker, C. N., Molpeceres, G., Rivilla, V. M., Majumdar, L., & Kastner, J. 2020, Astrophys. J., 897, 158.
Shingledecker, C. N., Álvarez Barcia, S., Korn, V. H., & Kästner, J. 2019, ApJ, 878, 80.
Shivani Misra A., Tandon P., 2017, Res. Astron. Astrophys., 17, 1.
Truhlar, D. G., ed. A. Kohen and H.-H. Limbach, Marcel Dekker, Inc., New York, 2006, p. 579.
Vazart, V., Calderini, D., Skouteris, D., Latouche, C. & Barone, V. 2015. J. Chem. Theory Comput. 11, 1165.
Vazart F., Latouche C., Skouteris D., Balucani N., Barone V., 2015, ApJ, 810, 111
Wynberg, A. 1989. Asymmetric autocatalysis: facts and fancy. J. Macromolecular Science, Part A, A26(8): 1033-1041.
Wonchoba, S. E., Hu, W.-P., Truhlar, D. G. In *Theoretical and Computational Approaches to Interface Phenomena*; Sellers, H. L., Golab, J. T., Eds.; Plenum Press: New York, 1994; p 1.
Wonchoba, S. E., Hu, W.-P.; Truhlar, D. G. 1995. *Phys. Rev. B. 51*, 9985.
Woon, D. E. 2002. Astrophys. J. 571, L177.
Zhang, X., Quan, D., Chang, Q., Herbst, E., Esimbek, J. & Webb, M. 2020. MNRAS. 497, 609.
Zaleski, D. P., Seifert, N. A., Steber, A. L., et al. 2013, ApJL, 765, L10.
Zuev, P. S., Sheridan, R. S., Albu, T. V., Truhlar, D. G. Hrovat, D. A. & Borden W. T. 2003. Science. 299, 867.